\documentclass[twocolumn,showpacs,preprintnumbers,amsmath,amssymb,pre]{revtex4}

\usepackage{graphicx}
\usepackage{dcolumn}
\usepackage{bm}

\begin{document}

\preprint{Phys. Rev. E {\bf 75}, 026217 (2007)}

\title{Phase-space structure of two-dimensional excitable localized structures}
\author{Dami\`a Gomila}
\email{damia@imedea.uib.es}
\author{Adrian Jacobo}
\email{jacobo@imedea.uib.es}
\author{Manuel A. Mat\'ias}
\email{manuel@imedea.uib.es}
\author{Pere Colet}
\email{pere@imedea.uib.es}
\affiliation{Instituto Mediterr\'aneo de Estudios Avanzados, IMEDEA (CSIC-UIB),
E-07122 Palma de Mallorca, Spain}
\homepage{http://www.imedea.uib.es/physdept}

\date{Published: February 28, 2007}

\begin{abstract}
In this work we characterize in detail the bifurcation leading to an excitable
regime mediated by localized structures in a dissipative nonlinear Kerr cavity
with a homogeneous pump. Here we show how the route can be understood through a
planar dynamical system in which a limit cycle becomes the homoclinic orbit of a
saddle point (saddle-loop bifurcation). The whole picture is unveiled, and the
mechanism by which this reduction occurs from the full infinite-dimensional
dynamical system is studied. Finally, it is shown that the bifurcation leads to
a excitability regime, under the application of suitable perturbations.
Excitability is an emergent property for this system, as it emerges from the
spatial dependence since the system does not exhibit any excitable behavior
locally.
\end{abstract}

\pacs{05.45.-a, 42.65.Sf, 89.75.Fb}

\maketitle

\section{Introduction} \label{sec:intro}

Localized structures (LS), or dissipative solitons, are spatio-temporal
structures that appear in certain dissipative media \cite{LSfirst}, and, in
particular, they have been found in a variety of systems, such as chemical
reactions \cite{LSPearson,LSLee}, gas discharges \cite{LSgasdis} or fluids
\cite{LSfirst}, among others. They are also found in optical cavities, due to 
the interplay of different effects, like diffraction, nonlinearity, driving, and
dissipation \cite{LSoptical,barland}. These structures, also known in this field
as cavity solitons, have to be distinguished from conservative solitons found,
for example, in propagation in fibers, for which there is a continuous family of
solutions depending, e.g., on the initial conditions. Instead, dissipative
solitons are unique once the parameters of the system have been
fixed. This fact makes this structures potentially useful in optical (i.e., fast
and spatially dense) storage and processing of information
\cite{barland,FirthOPN,Coullet04}.

Here we consider the dynamics LS in Kerr cavities, known as Kerr cavity
solitons, that arise as a consequence of a modulational (namely a
pattern-forming) instability of a homogeneous solution. In particular, they
exist in the parameter range where the homogeneous solution coexists with stable
subcritical (hexagonal) patterns. They share some properties with propagating
spatial (conservative) solitons in a Kerr medium, but there are interesting
differences. While in one transverse dimension (1D) Kerr spatial solitons are
stable, it is well known  that their 2D counterparts are unstable against
self-focusing collapse \cite{Sulem}. The stability and dynamics of $2$-D Kerr
cavity solitons are thus of particular interest, and their existence and
stability  has been studied in several papers
\cite{FirthLord,FirthPhSc,FirthJOSAB}.

Localized structures may develop instabilities like start moving, breathing or
oscillating. In the latter case, LS oscillate in time while remaining stationary
in space, like the oscillons found in a vibrated layer of sand \cite{oscillon}.
Oscillating LS are autonomous oscillons, and have been reported both in optical
\cite{FirthPhSc,FirthJOSAB,longhi} and chemical systems \cite{oscillonve}, and
appear when the LS exhibits a Hopf bifurcation.
In the present work we report on a route in which autonomous oscillating LS are
destroyed, leading to an excitability regime, extending upon the results
advanced in letter form in Ref. \cite{damia05}.

Typically a system is said to be excitable if while it sits at an stable fixed
point, perturbations beyond a certain threshold induce a large response before
coming back to the rest state. In phase space \cite{ErmenRinzel,Izhikevich}
excitability occurs for parameter regions where a stable fixed point is close to
a bifurcation in which an oscillation is created. Basically, there are two types
of excitability: one characterized by a response time (to come back to the fixed
point) within a relatively narrow range, also called Class II, and occurring in
the well known FitzHugh-Nagumo model, and also the case in which excitability is
mediated by a saddle point, also called Class I, and that exhibits an unbounded
distribution of response times. The route to excitability reported here
corresponds to the latter type of excitability. An interesting feature of this
system is that, while, typically, in excitable media excitability is also found
locally, i.e., in the zero-dimensional system, here we report a system in which
excitability is an emergent property: it is not present at the local level 
but it appears through a property of an spatio-temporal structure exhibited by
the system.

In the route reported in this work oscillating LS are made unstable in a global
bifurcation, namely a saddle-loop bifurcation, in which a limit cycle becomes
the homoclinic orbit of a saddle point. This bifurcation may occur generically
in $2$-variable continuous dynamical systems. Instead, the extended system
studied here lives in an infinite-dimensional phase space, and, moreover, does
not exhibit a spectrum with two slow modes that clearly dominate the dynamics, 
so an study is performed to show how the relevant dynamical behavior can be
reduced to a two-mode representation. This reduction is a common and powerful
procedure to study the dynamics of spatial systems exhibiting coherent
structures, however, the identification of the relevant modes is often highly
non trivial.

The plan of this paper is as follows. Fist of all, the model and overall
dynamical behavior exhibited by the system in parameter space are introduced in
Sections \ref{sec:model} and \ref{sec:overview}. Next, Section~\ref{sec:slbifn}
presents the case for the instability exhibited by LS through a saddle-loop
bifurcation, discussing the main evidences to support this conclusion.
Section~\ref{sec:modean} goes a step further in this direction, presenting a
more detailed study of how the dynamics of the system can be understood through
a simplified analysis, by performing an analysis in terms of modes. Finally,
some concluding remarks are given in Section~\ref{sec:conclu}.

\section{Model} \label{sec:model}

A prototype model describing an optical cavity filled up with a nonlinear Kerr
medium is the one introduced by Lugiato and Lefever \cite{LugiatoLefever87} with
the goal of studying pattern formation in this system. Later studies showed that
this model also exhibits LS in some parameter regions
\cite{FirthPhSc,FirthLord}.
The model, obtained through the mean-field approximation, describes the
dynamics of the slowly varying amplitude of the electromagnetic field
$E(\vec{x},t)$ in the paraxial limit, where $\vec{x}=(x,y)$ is the plane
transverse to the propagation direction $z$ on which the slow dynamics takes
place. The time evolution of the electric field can be, then, written as
\begin{eqnarray}
\frac{\partial E}{\partial t}=-(1+i \theta)E+i
\nabla^2E+E_0+i |E^2|E \label{kerreq}
\end{eqnarray}
after suitably rescaling the variables.

The first term in the rhs describes cavity losses (making the system
dissipative), $E_0$ is the homogeneous (plane wave) input field, $\theta$ the
cavity detuning with respect to $E_0$, and $\nabla^2=\partial^2/\partial
x^2+\partial^2/\partial y^2$ is the transverse Laplacian modeling the
diffraction. The sign of the cubic term indicates the so called self-focusing
case. Notice that in the absence of losses and an input field, the field can
be rescaled to $E \rightarrow E e^{i\theta t}$ to remove the detuning term and
Eq.~(\ref{kerreq}) becomes the Nonlinear Schr\"odinger Equation (NLSE). It is
well documented that in the NLSE in two spatial dimensions an initial condition
with enough energy collapses, namely energy accumulates at a point of space
leading to the divergence of the solution at a finite time \cite{collapse}.
Cavity losses prevent this collapse, although in the parameter region in which
localized structures are stable their dynamics are closely related to the
collapse regime. Eq.~(\ref{kerreq}) has a homogeneous steady state solution
$E_s=E_0/(1+(i(\theta-I_s))$, where $I_s=|E_s|^2$ \cite{LugiatoLefever87}. In
the following we use $I_s$, together with $\theta$, as our control parameters.

\section{Overview of the system behavior} \label{sec:overview}

\begin{figure}
\includegraphics{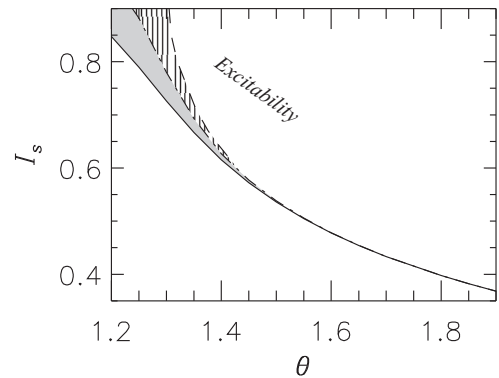}
\caption{Phase diagram of localized structures in the Kerr cavity. LS are
stable  in the shaded region and oscillate in the cross-hatched one (the
dot-dashed line between these two regions indicates a Hopf bifurcation). In the
lower part, below the saddle-node bifurcation (solid line), there are no LS,
while in the upper part, above the saddle-loop bifurcation (dashed line), the
system exhibits excitability.}
\label{fig:phasediag}
\end{figure}

Performing a two-parameter study of the system, it has been shown that a stable
regime of LS is found \cite{FirthJOSAB}. This
region of existence in the parameter space is shown in
Fig.~\ref{fig:phasediag}. This regime occurs for $I_s<1$, as at $I_s=1$ the
so-called modulation instability  takes place and the homogeneous solution
becomes unstable, leading to the formation of hexagonal patterns
\cite{LugiatoLefever87,Scroggie94}. For $I_s>1$ the homogeneous solution
continues to exist, although it is unstable. The hexagonal patterns are
subcritical, namely through a S-shape branch and, thus, they coexist with the
stable homogeneous solution for a certain parameter range.

\begin{figure}
\includegraphics[width=0.4\textwidth]{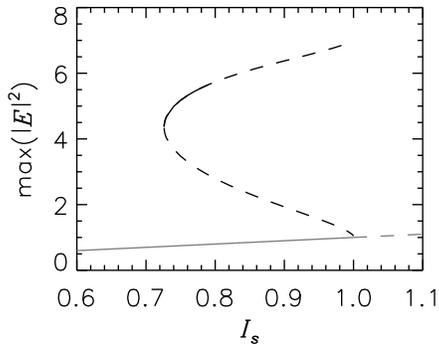}
\caption{Bifurcation diagram of stationary localized structures in the Kerr
cavity: max$(|E|^2)$ vs $I_s$ for $\theta=1.34$. Solid lines represent stable 
solutions and 
dashed lines unstable ones. The lowest branch corresponds to the homogeneous 
solution that becomes unstable at $I_s=1.0$. 
The upper and middle branches correspond to the stable and unstable LS,
respectively, and are originated at a saddle-node bifurcation. 
The upper branch becomes Hopf unstable for larger values of $I_s$.
\label{bifis0}} 
\end{figure}

This bistability regime is at the origin of the existence of stable LS, that
appear when suitable (localized) transient perturbations are applied. The LS can
be seen as a solution which connects a cell of the pattern with the homogeneous
solution. While the existence of LS in this bistable regime is quite generic in
extended systems, the stability of such LS strongly depends on the particular
system. The mechanism by which LS appear is a saddle-node (or
fold) bifurcation, as can be seen in Fig.~\ref{bifis0} for $\theta=1.34$ and 
$I_s\sim 0.655$ ($\vert E_0\vert^2\sim 4.5$) in which a pair of stable-unstable 
LS are created \cite{Champneys,Coullet00}.

\begin{figure}
\includegraphics[width=0.4\textwidth]{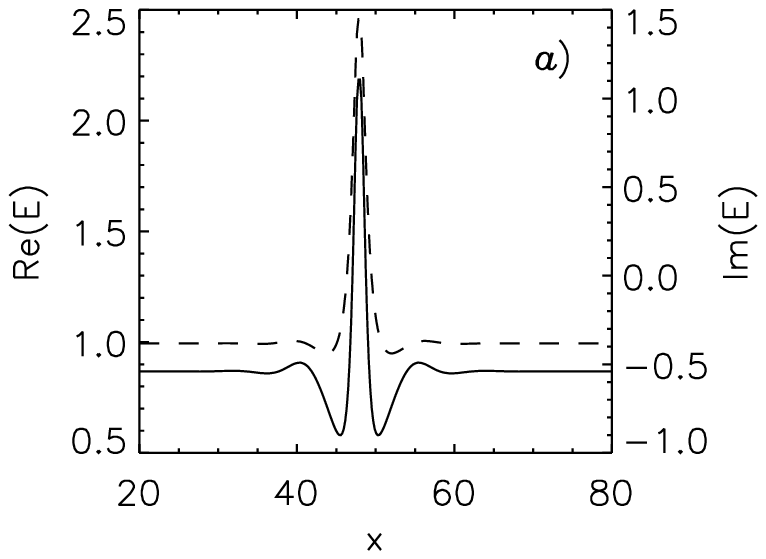}
\includegraphics[width=0.4\textwidth]{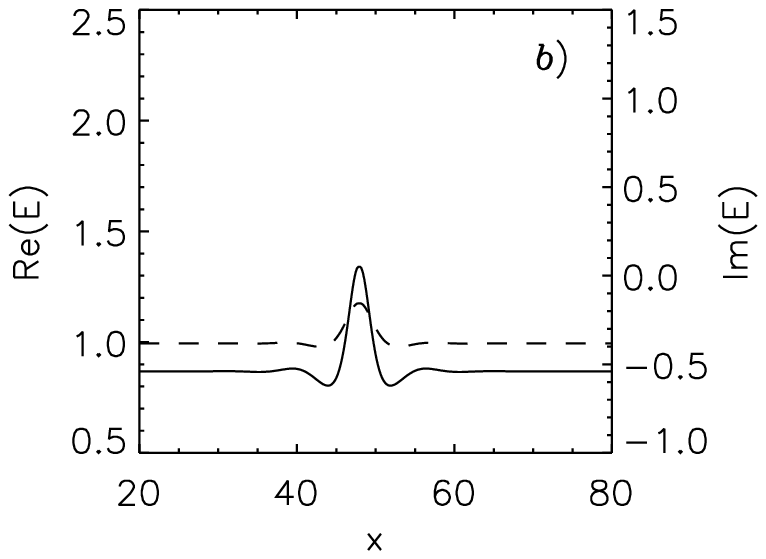}
\caption{Transverse cut of two LS, one from the upper (stable) branch 
a) and one from the middle (unstable) branch b) in 
Fig.~\ref{bifis0} ($I_s=0.9$). The solid (dashed) line corresponds to 
Re[$E$] (Im[$E$]).}
\label{stable_ls}
\end{figure}

The LS are rotationally symmetric around their center. Fig.~\ref{stable_ls}
shows a transverse cut of typical upper and middle branch LS. The upper branch
LS  remains stable for a range of values of $I_s$ and undergoes a Hopf
bifurcation leading to a limit cycle when $I_s$ is increased 
\cite{FirthLord,FirthJOSAB,Skryabin02}. The region in parameter space where 
LS oscillate is shown in Fig.~\ref{fig:phasediag}. Thus, in
these conditions a LS is an autonomous oscillon. An interesting connection to
the conservative case is that the growth of the LS during the oscillations
resembles the collapse regime observed for the $2D$ (or $2+1$) NLSE. In this
case, however, after some value is attained for the electric field, $E$,
dissipation arrests this growth.

As $I_s$ is further increased the amplitude of the limit cycle grows so that it
gets closer to the unstable (middle) branch structure. It is perhaps
surprising that the overall scenario can be understood qualitatively by
resorting to a planar dynamical system, i.e., one with a two-dimensional phase
space. These two phase space variables correspond to the amplitude of localized
modes of the system (this issue is studied in detail in
Section~\ref{sec:modean}). In the rest of the paper we will represent the main
features of the behavior exhibited by the system in terms of these
two-dimensional representation.

\begin{figure}
\includegraphics[width=0.4\textwidth]{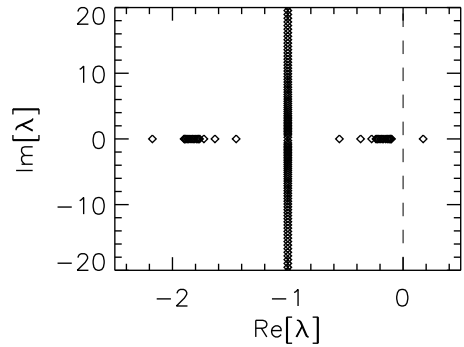}
\caption{Spectrum of the unstable (middle branch) LS for $\theta=1.304 785 92$
and $I_s=0.9$.}
\label{fig:spectrum}
\end{figure}

As shown in the Appendix, using a numerical method with arbitrary precision it 
is possible to determine the stability of the LS solutions. The spectrum of 
eigenvalues (of matrix ${\bf U}$, Eq.~(\ref{udef})) for an unstable (middle) 
branch LS is shown in Fig.~\ref{fig:spectrum}. There is only one positive 
eigenvalue so this structure has a single unstable direction in the full phase
space. In the reduced, planar phase space, it is a saddle point. Once it is
created, the middle branch LS does not undergo any bifurcation for the parameter
values explored in this paper and, so, remains a saddle point in phase space.
When the limit cycle (corresponding to the oscillating LS) touches the middle
branch the LS undergoes a so called \textit{saddle-loop} bifurcation, that is
the subject of Section~\ref{sec:slbifn}. Beyond this bifurcation an excitable
regime emerges, this regime will be described later in
Section~\ref{sec:excitability}.

\section{Saddle-loop bifurcation} \label{sec:slbifn}

A saddle-loop (also known as homoclinic or saddle-homoclinic) bifurcation 
is a global bifurcation in which a limit cycle becomes
biasymptotic to a (real) saddle point, or, in other terms, becomes the homoclinic orbit
of a saddle point (cf. \cite{Glendinningbook,Wiggins}), i.e., at criticality a
trajectory leaving the saddle point through the unstable manifold returns to it
through the stable manifold. Thus, at one side of this bifurcation one finds a
detached limit cycle (stable or unstable), while at the other side the cycle
does not exist any more, only its {\it ghost}, as the bifurcation creates an
exit slit that makes the system dynamics to leave the region in phase space
previously occupied by the cycle. Thus, after the bifurcation the system
dynamics jumps to another available attractor. In the present case this
alternative attractor is the homogeneous solution.

Let us take $\theta$ as the control parameter and assume that the saddle-loop
bifurcation occurs for $\theta=\theta_{SL}$, and, for convenience, and without
lack of generality, let us assume that $\theta<\theta_{SL}$ corresponds to the
oscillatory side, where the limit cycle is detached of the saddle point, while,
in turn, $\theta>\theta_{SL}$ corresponds to the side where one only has a fixed
point solution.
The fact that the bifurcation is global, implies that it cannot be detected
locally (a local eigenvalue passing through zero), but one can still resort to
the Poincar\'e map technique \footnote{
The Poincar\'e map can be constructed through two cross
sections, i.e., two planes that are transversal to the limit cycle, and that are
placed slightly before and after the closest approach of the cycle to the saddle
point. From these two planes one can construct two maps: the so-called local (or
linear or singular) map, $T_0$, that takes the flow from the plane before the
saddle point to the plane after the saddle point, and it is dominated by the
saddle point, and the global (or nonlinear) map, $T_1$, that takes the flow all
the way from the plane past the saddle point through all the limit cycle back to
the plane before the saddle point. The complete Poincar\'e map is the
composition of these two maps. It has to be remarked that the $T_0$ map is
unbounded, as the return time is infinity at the onset of the global
bifurcation.}
to analyze it, and, interestingly, the main
features of the bifurcation can be understood from the knowledge of the linear
eigenvalues of the saddle. 

The case studied here is the simplest: a saddle point with real eigenvalues, say
$\lambda_s<0$ and $\lambda_u>0$, in a $2$-dimensional phase space. Strictly
speaking, in our case the saddle has an infinite number of eigenvalues
(Fig.~\ref{fig:spectrum}), but only two eigenmodes take part in the dynamics
close to the saddle. This will be studied in more detail in
Sec.~\ref{sec:modean}.
It is convenient to define the so called {\it saddle index\/}
$\nu=-\lambda_s/\lambda_u$ and {\it saddle quantity\/}
$\sigma=\lambda_s+\lambda_u$.
It can be shown \footnote{For the details refer, e.g., to Sec.
12.3 of Ref. \cite{Glendinningbook}} that for $\sigma<0$, or $\nu>1$, at the
side of the saddle-loop bifurcation where one has a detached cycle, this
cycle is stable, while for $\sigma>0$ ($\nu<1$), the cycle is unstable.
Analogously, one can study the period of the cycle close to this bifurcation,
and to leading order it is given by \cite{GaspardSL},
\begin{equation}
T\propto -\frac{1}{\lambda_u} \ln\vert \theta-\theta_{SL}\vert\ .
\label{scalingsl}
\end{equation}
This expression is accurate for $\theta$ close enough to $\theta_{SL}$.
Interestingly, the transient times spent by a trajectory in the ghost region
after the cycle has ceased to exist, close enough to the bifurcation point,
also show this scaling.

\begin{figure}
\includegraphics[width=0.35\textwidth]{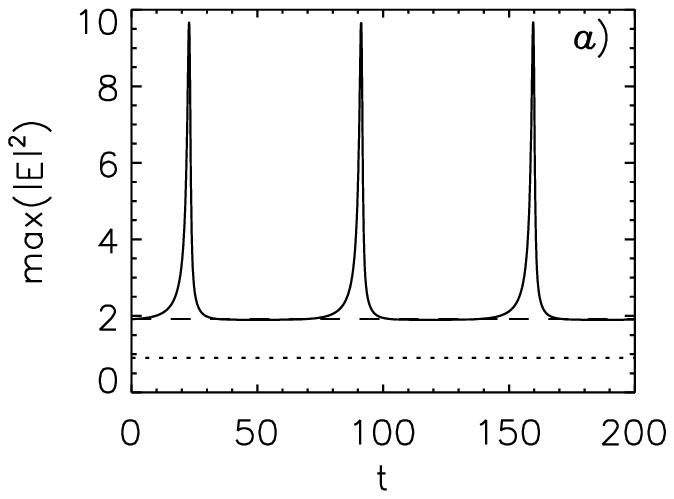}
\includegraphics[width=0.35\textwidth]{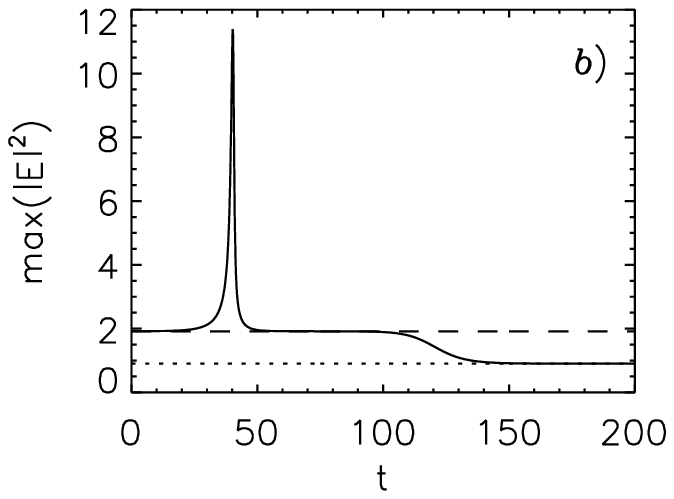}
\caption{Maximum intensity of the LS as function of time for $I_s=0.9$.
a) Oscillatory trajectory for $\theta=1.3047859$ (just below  
$\theta_{SL}$). b) Excitable trajectory starting from an initial condition 
very close but above the saddle point in the phase space ($\theta=1.3047860$,
just above $\theta_{SL}$)}
\label{fig:traj}
\end{figure}

\begin{figure}
\includegraphics[width=0.4\textwidth]{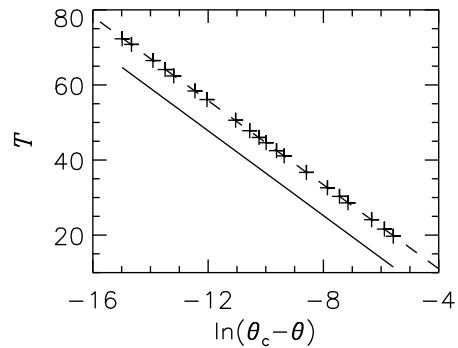}
\caption{Scaling of the period in the saddle-loop bifurcation. Crosses
correspond to numerical simulations while the solid line, arbitrarily
positioned, has a slope $1/\lambda_u$ with $\lambda_u=0.177$ obtained from the
stability analysis of the unstable LS.}
\label{fig:homoscaling}
\end{figure}

From a numerical viewpoint, we will characterize the occurrence of a saddle-loop
bifurcation in the system by studying the scaling of the period of the
oscillations. The bifurcation point will be characterized by the fact that
approaching from the oscillatory side the period diverges to infinity, and also
because past this bifurcation point the LS disappears and the system relaxes to
the homogeneous solution as shown in Fig.~\ref{fig:traj} for $I_s=0.9$. For 
this value of $I_s$ the saddle-loop takes place at $\theta_{SL}=1.30478592$. In
the figure the time evolution of the maximum of the LS is plotted for two values
of the detuning differing in $10^{-7}$, one just above and the other just below 
$\theta_{SL}$. Figure~\ref{fig:homoscaling} contains a
logarithmic-linear plot of the period versus a control parameter, that exhibits,
as expected, a linear slope. Furthermore, one can confront the value of the
slope obtained form the simulations with its theoretical prediction, Eq.
(\ref{scalingsl}), namely $1/\lambda_u$. The full spectrum of the middle branch
soliton for $\theta=\theta_{SL}$ (calculated as described in the Appendix) is 
shown in Fig.~\ref{fig:spectrum}. The agreement between the simulations and 
theoretical slopes is within 1\%.

A comment is in place here regarding the spectrum shown in
Fig.~\ref{fig:spectrum}.  The spectrum is formed by a stable continuous
(although numerically discretized) and
a discrete spectrum with a positive ($\lambda_u=0.177$) and a negative
($\lambda_s=-2.177$) eigenvalue. Having this spectrum in mind is perhaps
surprising that one can describe the bifurcation route very well qualitatively,
and to some extent quantitatively (cf. the observed scaling law, 
Fig.~\ref{fig:homoscaling}), resorting to a
planar dynamical system when many modes could be, in principle, involved. The
first mode of the planar theory univocally corresponds to the positive
(unstable) eigenvalue, $\lambda_u=0.177$, while, in first approximation, the
second mode should correspond to the the second, closest to zero, eigenvalue.
This eigenvalue belongs however to a continuum band and the arbitrarily close
eigenvalues of its band could play a role in the dynamics, modifying the planar
theory. Moreover, considering this mode $\lambda \sim -0.10$ the saddle-index
$\nu=-\lambda/\lambda_u <1$ indicating that the cycle emerging from the
saddle-loop should be unstable, although we observe otherwise. The analysis of the
modes of the unstable LS and dimensionality of the phase space is addressed in
detail in the next section.

\section{Mode Analysis} \label{sec:modean}

\begin{figure}
\includegraphics[width=0.4\textwidth]{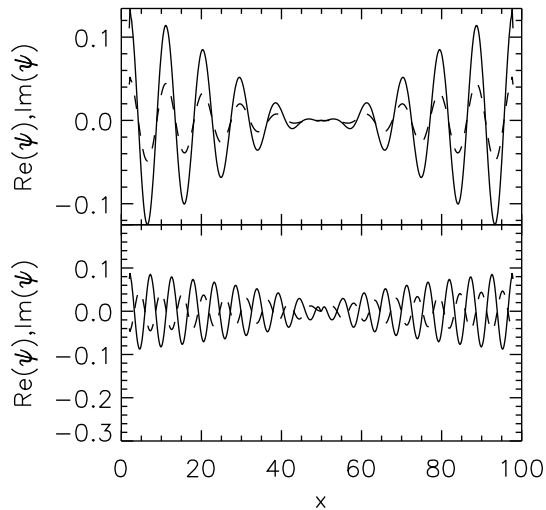}
\caption{Stable extended modes from the continuous band. The top (bottom) panel
shows the transverse cut of the mode associated to the eigenvalue $\lambda=-0.1$
($\lambda=-1+i0.24$) of Figure~\ref{fig:spectrum}.The solid (dashed) line
indicates the real (imaginary) parts of
the eigenmode.}
\label{fig:panel2}
\end{figure}

\begin{figure}
\includegraphics[width=0.4\textwidth]{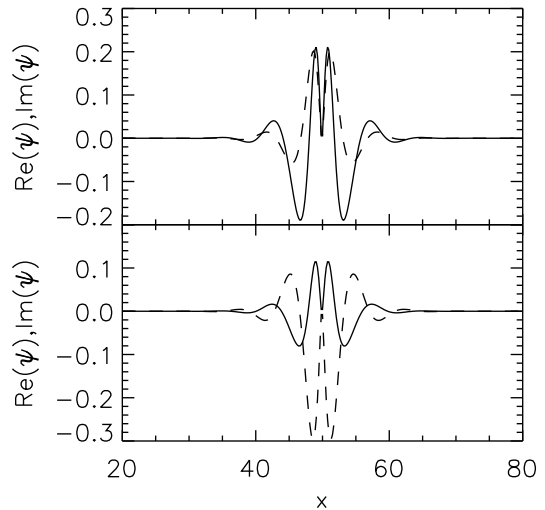}
\caption{Transverse cut of the unstable (top) and the most stable (bottom)
modes of the unstable LS. These modes are associated to the eigenvalues
$\lambda_u=0.177$ and $\lambda_s=-2.177$ of Fig.~\ref{fig:spectrum}
respectively. The solid (dashed) line indicates the real (imaginary) parts of
the eigenmode.}
\label{fig:panel1}
\end{figure}

In this section we analyze the dynamics in terms of the
modes obtained in performing the stability analysis of the middle branch LS in 
a parameter region close to the saddle-loop bifurcation, as described in  
Appendix A. By plotting the spatial profile of the modes one obtains a clue  to
identify the relevant modes for the dynamics. It turns out that most of the
modes of the stable spectrum are  delocalized. Figure~\ref{fig:panel2} contains
a representation of  two such delocalized modes. The bands of extended modes 
correspond to modes of the homogeneous background, and are, except for
a radial dependence coming from the fact that we are using radial instead of
Cartesian coordinates, basically Fourier modes. The main difference between these
modes is the wavenumber of their oscillations (see
Fig.~\ref{fig:panel2}). There are however two exceptions: two localized modes
which are the one  corresponding to the unstable direction and the most stable
mode,  namely that with eigenvalue $\lambda_s=-2.177$. The spatial profile of
these two modes is shown in Figure~\ref{fig:panel1}. Since the dynamics of the
LS remains localized in the space, this is an  indication that only these two
localized modes take part in the dynamics. To check this hypothesis we have
projected the two trajectories shown in  Fig.~\ref{fig:traj} for parameters
close to the saddle-loop bifurcation onto all the eigenmodes of the unstable LS
and observed that only the two localized modes have a significant amplitude.

From the knowledge of the spectrum and the relevant eigenmodes, we can now
explain the stability of the orbits emerging out of the bifurcation, namely
through the saddle index introduced above. Computing this index for the two
modes that participate in the saddle-loop bifurcation one obtains
$\nu=2.177/0.177>1$, what fits perfectly with the fact that the cycle that
detaches at one side of the bifurcation point is stable. Thus, one may
understand that all the dynamical instability scenario of the LS can be analyzed
qualitatively in a planar dynamical system.

\begin{figure}
\includegraphics[width=0.4\textwidth]{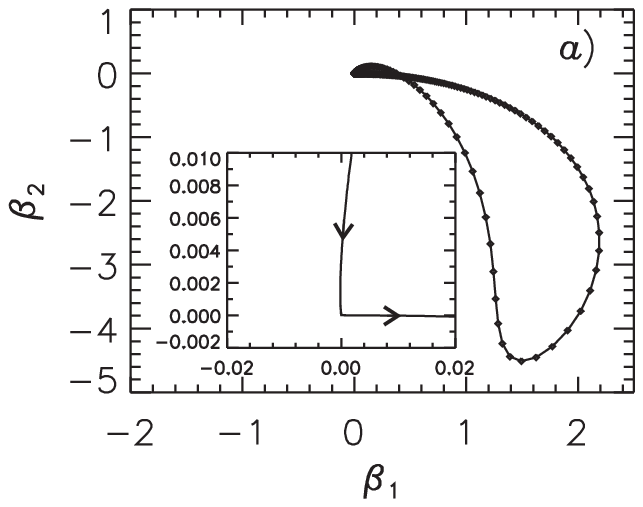}
\includegraphics[width=0.4\textwidth]{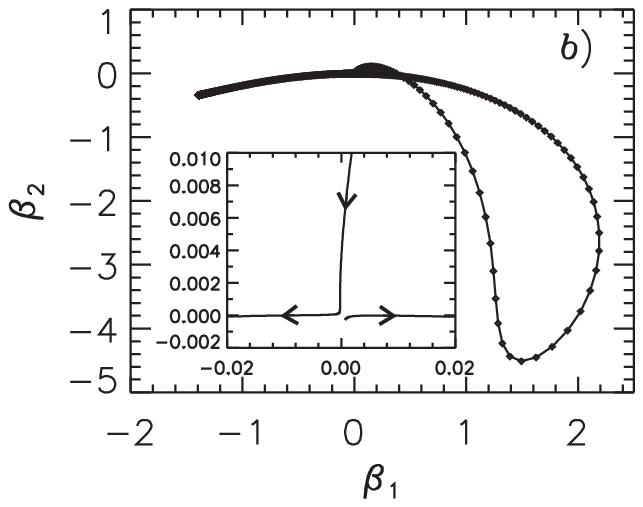}
\caption{Reconstructed phase space by finding the amplitude of the deviation of
the trajectory from the unstable LS in the unstable ($\beta_1$) and the
most stable ($\beta_2$) modes of the unstable LS.
Panel a) corresponds to an oscillatory trajectory while panel b) to an
excitable one. The symbols are equispaced in time
along the trajectory, so sparse symbols indicate fast dynamics while dense
symbols indicate slow dynamics. The saddle point is at $(0,0)$. The inset is a
close up of the linear region around the saddle.}
\label{figqpp1}
\end{figure}

A better understanding of the dynamical route, and a justification of the role
of the two participating localized modes, stable and unstable, can be obtained
through a closer scrutiny of the {\it linear\/} region, namely the region close
to the saddle point (or alternatively, the region defined by the singular map,
or close to it). Figure~\ref{fig:traj}a contains a time trace of one such
trajectory in the region in which the limit cycle is stable, but close to the
saddle-loop bifurcation. 
Following Appendix A we project the deviation of the trajectory from the
unstable LS (saddle point) onto the most stable and the unstable
eigenvectors of the adjoint Jacobian matrix of the unstable LS.
These projections are the amplitudes of the unstable ($\beta_1$) and the
most stable ($\beta_2$) modes of the unstable LS (modes whose profile is
shown in Fig. 8). The trajectory enters
the linear region through the stable mode and leaves the region through the 
unstable one. This behavior is clear in the insets of Figure~\ref{figqpp1}.
Next, we reconstruct the qualitative sketch of the bifurcation shown in Fig.~2
of \cite{damia05} from the knowledge of the projections onto the modes, i.e. we
represent the trajectories before and after the saddle-loop bifurcation in mode
space. Thus, Figure \ref{figqpp1} contains a quantitative, reconstructed,
$2$-dimensional phase space from the two localized modes involved in the
transition for a set of parameter values in the a) oscillatory and  b) excitable
side of the transition. Close to the saddle, the linear dynamics takes place
on a plane, but away from this point the nonlinear dynamics bends the trajectory
out of the plane into the higher dimensional space, hence, the apparent 
crossing of the trajectory in Fig.~\ref{figqpp1}.

This is the final numerical confirmation that the infinite-dimensional
dynamical system on which LS live can be reduced to an excellent degree
of precision to a $2$-dimensional dynamical system, and that the
picture is fully consistent with a saddle-loop bifurcation.

\section{Excitable behavior} \label{sec:excitability}

As in our case the saddle-loop bifurcation involves a fixed point (the 
homogeneous solution), 
on one side of the bifurcation, and an oscillation, on the other, the
system is a candidate to exhibit excitability \cite{Izhikevich}. It must be
stressed that excitable behavior is not guaranteed {\it per se\/} after a
saddle-loop bifurcation, and, in particular one needs a fixed point attractor
that is close enough to the saddle point that destroys the oscillation. The
excitability threshold in this type of system is the stable manifold of the
saddle point, what implies that the observed behavior is formally Class I
Excitability \cite{Izhikevich}, i.e., the excitability is characterized by
response times that can be infinite (if a perturbation hits exactly the stable
manifold of the fixed point), or, conversely, frequencies starting from zero.
In our system, the excitable threshold reduces by increasing $I_s$
(Fig.~\ref{bifis0}), since the middle branch LS (the saddle point)
gets progressively closer to the homogeneous solution (fixed point).

\begin{figure}
\includegraphics[width=0.3\textwidth]{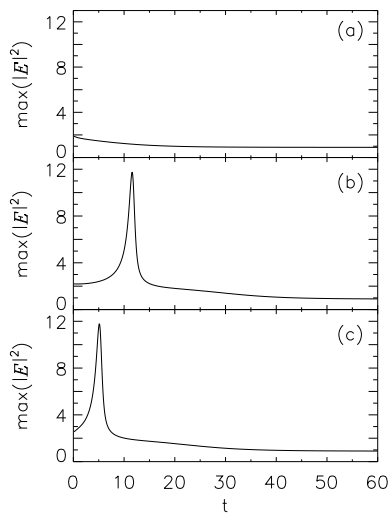}
\caption{Time evolution of the maximum intensity starting from the 
homogeneous solution ($I_s=0.9$) plus a localized perturbation of the form of 
the unstable LS multiplied by a factor $0.8$ (a), $1.01$ (b) and $1.2$ (c).
}
\label{fig:exctraj} 
\end{figure}

\begin{figure}
\includegraphics[width=0.4\textwidth]{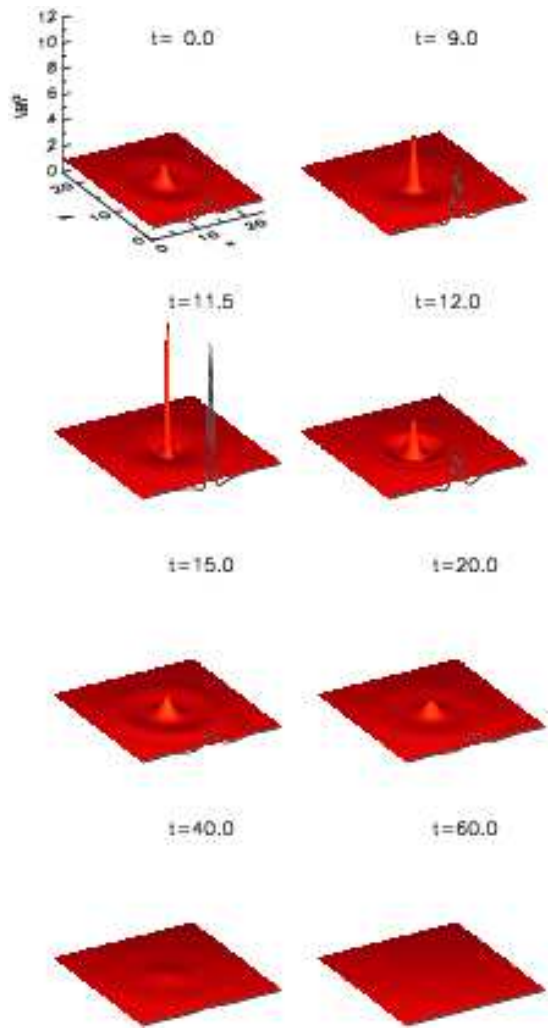}
\caption{(Color online) $3D$ plots showing the transverse intensity profile at
different times for the trajectory shown in Fig.~\ref{fig:exctraj}(b). The solid
lines show a cut of the structure through the center.}
\label{fig:exc3D} 
\end{figure}

This excitability scenario was shown in Ref. \cite{damia05}, and in  parameter
space it is found in the region above the dashed line corresponding  to the
saddle-loop bifurcation shown in Fig.~\ref{fig:phasediag}. 
Fig.~\ref{fig:exctraj} shows the resulting trajectories after applying a 
localized  perturbation in the direction of the unstable LS with three
different  amplitudes: one below the excitability threshold (a), and two above: 
one very close to threshold (b) and the other well above (c).  For the
below-threshold perturbation the system decays exponentially to the  homogeneous
solution, while for the above-threshold perturbations a long excursion in phase
space is performed before returning to the stable fixed point. The refractory
period for the perturbation just above the excitability threshold is appreciably
longer due to the effect of the saddle.  The spatio-temporal dynamics of the
excitable localized structure is shown in  Fig.~\ref{fig:exc3D}. After an
initial localized excitation is applied,  the peak grows to a large value until
the losses stop it. Then it decays exponentially until it disappears. A remnant
wave is emitted out of the center dissipating the remaining energy.

At this point it is worth noting that neglecting the spatial dependence 
Eq.~(\ref{kerreq}) does not present any kind of excitability. The excitable
behavior is an emergent property of the spatial dependence and it is strictly 
related to the dynamics of the 2D LS. Without spatial
dependence, excitability as a result of a saddle-loop bifurcation has been
observed in different systems \cite{Plaza,ErmenRinzel,Izhikevich}.

Finally, it is interesting to remark that the excitable region in parameter
space is quite large (cf. Fig.~\ref{fig:phasediag}) and, therefore,  potentially
easy to observe experimentally. While this excitable behavior belongs to Class I
(the period diverges to infinity  when a perturbation hits the saddle), due to
the logarithmic scaling law for the period (\ref{scalingsl}), the parameter
range over which the period  increases dramatically is extremely narrow (cf.
Fig. 3(a) in Ref. \cite{damia05}). Therefore, from an operational point of view, 
systems exhibiting this scenario might not be classified as Class I
excitable,  as the large period responses may be easily missed
\cite{IzhikevichDSN}.

\section{Takens-Bogdanov Point} \label{sec:TBpoint}

The saddle-loop (or homoclinic) bifurcation is, in some sense, not {\it
generic\/}. Namely, that a tangency between a limit cycle and a saddle point
occurs exactly such that it happens simultaneously at both the sides of the
stable and unstable manifolds is, in principle, not to be expected generically.
In fact, also due to the fact that global bifurcations are not always easy to
detect, showing that a dynamical system exhibits a certain type of
codimension-$2$ point is the most convincing argument for the existence of such
bifurcations.

An scenario in which the unfolding of a codimension-$2$ point yields
a saddle-loop (or homoclinic) bifurcation is a Takens-Bogdanov (TB) point
\cite{GuckenHolmes,Kuznetsov}.
Namely, a double-zero bifurcation point in which a saddle-node bifurcation
line and the zero-frequency limit of a Hopf bifurcation line (thus, no longer a
Hopf line in the crossing point) meet in a two-parameter plane. The particular 
feature that, at the
TB point, the Hopf line has zero frequency allows this codimension-$2$ 
bifurcation to occur in a two-dimensional phase space. This bifurcation
has to be distinguished from the occurrence of a crossing between a saddle-node 
and a Hopf lines at non-zero frequency, known as Gavrilov-Guckenheimer (saddle 
node-Hopf point), that requires a three-dimensional phase space to take
place. One can prove that from the unfolding of a TB point a
saddle-loop line, apart from the saddle-node and Hopf lines, emerges
\cite{GuckenHolmes,Kuznetsov} from the TB point.

\begin{figure}
\includegraphics[width=0.4\textwidth]{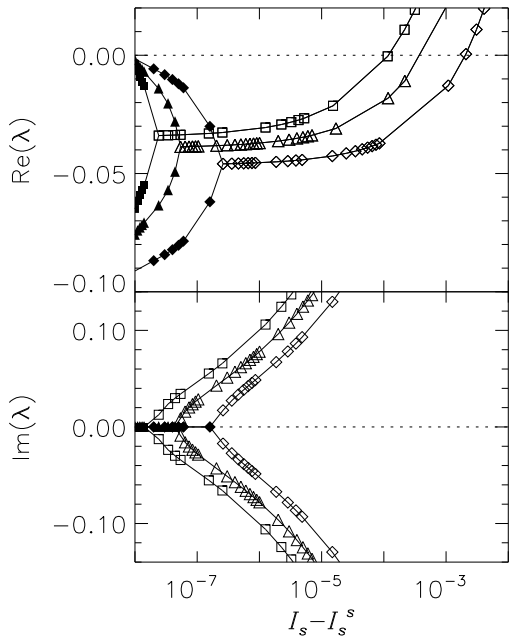}
\caption{Real part (upper panel) and imaginary part (lower panel) of the
eigenvalues of the upper branch LS for three vertical cuts in Fig.
\ref{fig:phasediag} corresponding to three different values  
$\theta$: squares, $1.7$; triangles, $1.5$; rhombs, $1.4$ versus the 
difference between $I_s$ and its value at the saddle-node bifurcation, 
$I_s^s(\theta)$.}
\label{fig:TB}
\end{figure}

This can be checked in Fig.~\ref{fig:phasediag}, in which a two-parameter
bifurcation plot is presented as a function of the two parameters of the system:
$I_s$ and $\theta$. The problem here is that the saddle-node and Hopf lines tend
to meet only asymptotically, namely when $\theta\rightarrow\infty$. In Ref.
\cite{damia05} we checked already that the distance between the saddle-node and
the Hopf lines decreases as one increases $\theta$ (the same happens with the
saddle-loop line). By calculating the eigenvalues, it can be seen that, indeed
the frequency (viz. their imaginary part) goes to zero as one approaches the TB
point. Fig.~\ref{fig:TB} displays the two eigenvalues with largest real part of
the upper branch LS for parameter values corresponding to three vertical cuts
of  Fig.~\ref{fig:phasediag}. Open symbols correspond to eigenvalues with a 
non-zero imaginary part while filled symbols are associated to real
eigenvalues.  Where the open symbols cross zero in the upper panel of Fig.
\ref{fig:TB} signals the Hopf bifurcation while where the filled symbols cross
zero signals  the saddle-node bifurcation. The origin for the three plots is
taken at the  saddle-node bifurcation. At some point along the branch of the two
complex conjugate eigenvalues  associated to the Hopf bifurcation the imaginary
part vanishes leading to two branches of real eigenvalues, the largest of which
is precisely the  responsible of the saddle-node bifurcation. As detuning
increases the Hopf and  saddle-node bifurcation points gets closer and closer
but the structure of  eigenvalues remains unchanged so that when the Hopf and
saddle-node  bifurcation will finally meet the Hopf bifurcation will have zero
frequency,  signaling a TB point.

The TB point takes place asymptotically in the limit of large detuning $\theta$ 
and small pump $E_0$. In this limit Eq.~(\ref{kerreq}) becomes the 
conservative NLSE \cite{FirthLord}. 
The Hopf instability in this limit was studied in \cite{Skryabin02}, 
where evidence of the double-zero bifurcation point was given, however the 
unfolding leading to the scenario presented here was not analyzed.

\section{Concluding remarks} \label{sec:conclu}

In this work a detailed study of the instabilities of LS solutions in
homogeneously pumped nonlinear Kerr cavities and the associated excitability
route  first reported in Ref. \cite{damia05} is carried out. In that study, it
was shown that the instability that leads to the destruction of oscillatory LS
found in this system can be characterized by a saddle-loop (homoclinic)
bifurcation, in which, in phase space, the oscillation (a limit cycle) becomes
the homoclinic orbit of a saddle point. This scheme is able to explain
accurately quantitative aspects of the transition, like scaling law for the
divergence of the period of the oscillation at the bifurcation point.

After a close scrutiny there is at least an aspect that may sound puzzling in 
this picture: the system under study is described by a $2$-D nonlinear Partial 
Differential Equation, with an infinite-dimensional phase space. Instead, the 
reported saddle-loop bifurcation minimally needs a $2$-D dimensional system with
a limit cycle and a saddle point, what is coherent with the fact that the
bifurcation {\it is born\/} at a Takens-Bogdanov codimension-$2$ point. One can
devise a kind of slaving principle, in which the slowest modes (with the
closest eigenvalues to zero) dominate the slow dynamics. However, the two
leading  eigenvalues of the saddle point close to bifurcation do not explain
the scenario (give the wrong stability for the limit cycle). For all this, we
have engaged in showing the reason why such a $2$-dimensional reduction is 
successful in explaining the dynamics, and, in particular, what happens with the
stability of the emerging limit cycle.

The main result of this paper is that it is possible to recover quantitatively,
from the full system, the qualitative $2$-dimensional sketch of the saddle-loop
bifurcation, in which two modes participate: the single unstable mode, and a
kind of conjugate stable mode (buried in the sea of stable modes), with the
property that both modes are the only two localized modes of the system.

Apart from this kind of fundamental result, the main interest of the present 
work is to prove excitability in an extended system as an emerging property,
{\it i.e.}, not present locally in the spatio-temporal system, but emerging 
through one of its solutions. Excitability is possible in classes of systems
in which an oscillation is destroyed at a bifurcation yielding, at the other
side, a fixed point solution. In the present scenario, mediated by a saddle-loop
bifurcation, excitability is not generic, and requires the availability of
a close enough fixed point solution: the homogeneous solution, in
the present case.

\begin{acknowledgments}
We thank D. Paz\'o for useful discussions.
We acknowledge financial support from MEC (Spain) and FEDER:
Grants FIS2004-00953 (CONOCE2), FIS2004-05073-C04-03,
FIS2006-09966, and 
TEC2006-10009 (PhoDeCC). AJ acknowledges financial support from MEC.
\end{acknowledgments}

\section*{Appendix A}

In this appendix we describe in some detail the numerical methods used 
throughout this paper.

For numerical simulations, we integrate Eq.~(\ref{kerreq}) using a
pseudo-spectral method where the linear terms in Fourier space are integrated
exactly while the nonlinear ones are integrated using a second-order in time
approximation \cite{Montagne}. Periodic boundary conditions are used, since 
they are convenient for the pseudospectral code. The system size is large 
enough to ensure that the electric field reaches the homogeneous steady state 
well before the boundaries. A square lattice of size $512\times 512$ points  was
used. The space discretization was taken $dx=0.1875$ while the time step was
$dt=10^{-3}$

To study the stability of the stationary LS solutions of Eq.~(\ref{kerreq}), we
set $E = E_s\,(1 + A)$, so that $A(x,y)$ describes the 
solution without the homogeneous background,
\begin{eqnarray}
\frac{\partial A}{\partial t} &=& -(1+i \theta) A +
i\nabla^2 A \nonumber \\
&&+i I_s \left(2A+A^{\star}+A^2+2\vert A\vert^2+
\vert A\vert^2\,A\ \right),
\label{eq:kerreqa}
\end{eqnarray}
where Eq. (\ref{eq:kerreqa}) is obtained directly from Eq. (\ref{kerreq}),
without any approximation. To obtain the stationary solutions one may
numerically solve the rhs of Eq. (\ref{eq:kerreqa}) equated to zero. However, 
since we have 2 spatial dimensions and the self-focusing dynamics involve very
large wavenumbers with very fast dynamics, this is a difficult and time 
consuming task.
Instead, we can take advantage of the fact that the LS structures are
rotationally symmetric with  respect to their center, so that they can be
described in terms of the 1D  radial equation for $A(r)$
\begin{eqnarray}
\frac{\partial A}{\partial t}&=& -(1+i \theta) A +i\left(\frac{\partial^2}
{\partial^2 r} +\frac{1}{r}\frac{\partial}{\partial r}\right) A \nonumber \\
&&+i I_s \left(2A+A^{\star}+A^2+2\vert A\vert^2+
\vert A\vert^2\,A\ \right).
\label{eq:kerrradial}
\end{eqnarray}
Steady state LS solutions for this system, both stable and unstable, are found
by equating to zero the lhs of Eq. (\ref{eq:kerrradial}). The boundary 
conditions for this problem are such that the derivatives are zero at the 
boundaries, i.e.,
$\partial A/\partial r (r=0)=\partial A/\partial r (r=L)=0$, where the 
system size $L$ is large enough to ensure that the electric field approaches 
smoothly the homogeneous solution ($A(r) \rightarrow 0$) before reaching the 
boundary. 

The stability of  steady state LS against radial and azimuthal perturbations is 
obtained, cf. Ref. \cite{McSloy02}, by linearizing Eq. (\ref{eq:kerreqa}) around 
the corresponding, numerically obtained, stationary solution $A_{LS}$. This 
yields a linearized equation for the time evolution of the perturbations 
$\delta A(r,\phi,t)=A(r,\phi,t)-A_{LS}(r)$. 
The solutions of the linear problem can be written as
\begin{equation}
\delta A=[R_+(r)\,e^{i m\phi}+R_-(r)\,e^{-i m\phi}] 
\exp(\lambda\,t)\ ,
\label{varA}
\end{equation}
where $m$ is the wavenumber of the azimuthal perturbation. This yields the 
eigenvalue problem
\begin{equation}
 {\bf U} {\bf\Psi} = \lambda {\bf\Psi}
\label{eigenvalue_problem}
\end{equation}
where ${\bf\Psi}=(R_+,R_-^\star)^{\top}$ and ${\bf U}=\left( \begin{array}{cc}
U_+ & U_- \\  U_-^\dagger & U_+^\dagger \end{array} \right)$, with 
\begin{eqnarray}
U_+&=&-(1+i\theta)+i\left(\frac{\partial^2}{\partial^2 r}
+\frac{1}{r}\frac{\partial}{\partial r}-\frac{m^2}{r^2}
\right) \nonumber \\
&& + i 2 I_s \left( 1+A_{LS}+A_{LS}^*+\vert A_{LS}\vert^2 \right) \nonumber \\
U_-&=&i I_s \left( 1+2A_{LS}+ A_{LS}^2 \right), 
\label{udef}
\end{eqnarray}
is the Jacobian. For purely radial perturbations ($m=0$) $R_-=R_+$.
The matrix ${\bf U}$ is time-independent as it is evaluated at the 
stationary LS (stable or unstable) under study.

The problem reduces, thus, to finding the eigenvalues, $\lambda$, and
eigenvectors, ${\bf \Psi}$, where it is important to mention that ${\bf U}$ is a
complex matrix and, thus, the eigenvectors are complex quantities in general.
Due to the symmetry of $\textbf{U}$ the eigenvalues are either real or pairs of
complex conjugates. This last property stems from the fact that, considering 
the real and imaginary parts of $A_{LS}$, ${\bf U}$ can be rewritten as a real
matrix. We note also that, due to the discretization of the space, ${\bf \Psi}$
becomes a vector whose dimension is $2N$. The set of eigenvectors ${\bf \Psi}_i$
($i=1,2N$) form a basis, and their amplitudes define a natural phase space where
studying the dynamics of LS. 

However, $\textbf{U}$ is not a self-adjoint operator so, the set of eigenmodes
does not form an orthogonal basis. To find the components of a field profile on
a mode ${\bf \Psi}_i$ one has to project it onto the corresponding eigenmode
${\bf  \Phi}_i$ of the adjoint Jacobian matrix ${\bf U^\dagger}$. In
Section~\ref{sec:modean} we are interested in the deviation of the field profile
from the unstable LS (saddle point) ${\bf \delta A }= (\delta A,\delta
A^\star)^{\top}$ . In particular we calculate the components of this deviation
on the unstable and the most stable eigenmodes of the Jacobian matrix of the
unstable LS. These amplitudes are given by  $\beta_i=\int {\bf \Phi}_i^{\top
\star} \cdot {\bf \delta A} dr$ ($i=1,2$), where $\Phi_1$ is the unstable and
$\Phi_2$ the most stable eigenmodes of the adjoint Jacobian matrix.

We should note that in the work reported here the LS are always inside the 
region in which they are stable versus azimuthal perturbations 
\cite{FirthJOSAB}, so all the instabilities described in the text are 
obtained for $m=0$. As final comment, the stability problem of stationary LS, 
that in principle live in an infinite-dimensional phase space, is reduced 
numerically to a stability problem in a finite-dimensional, 
albeit large, phase space.

\end{document}